\documentclass[useAMS,usenatbib,usegraphicx,referee]{mn2e}

\begin{document}

\title[Gas scale heights in four dwarf galaxies]{Theoretical determination of H\,{\Large\bf I} vertical scale heights in the dwarf galaxies: DDO\,154, Ho\,II, IC\,2574 \& NGC\,2366}
\author[A. Banerjee et al.]{Arunima Banerjee$^1$, Chanda J. Jog$^1$, Elias Brinks$^2$ and Ioannis Bagetakos$^2$\\
$^1$Department of Physics, Indian Institute of Science, Bangalore 560012,India\\
$^2$Centre for Astrophysics Research, University of Hertfordshire, College Lane, Hatfield AL10 9AB, United Kingdom}

\maketitle

\begin{abstract}
In this paper, we model dwarf galaxies as a two--component system of gravitationally coupled stars and atomic hydrogen gas in the external force field of a pseudo--isothermal dark matter halo, and numerically obtain the radial distribution of \mbox{H\,{\sc i}} vertical scale heights. This is done for a group of four dwarf galaxies (DDO\,154, Ho\,II, IC\,2574 and NGC\,2366) for which most necessary input parameters are available from observations. The formulation of the equations takes into account the rising rotation curves generally observed in dwarf galaxies. The inclusion of self--gravity of the gas into the model at par with that of the stars results in scale heights that are smaller than what was obtained by previous authors. This is important as the gas scale height is often used for deriving other physical quantities. The inclusion of gas self--gravity is particularly relevant in the case of dwarf galaxies where the gas  cannot be considered a minor perturbation to the mass distribution of the stars. We find that three out of four galaxies studied show a flaring of their \mbox{H\,{\sc i}} disks with increasing radius, by a factor of a few within several disk scale lengths. The fourth galaxy has a thick  \mbox{H\,{\sc i}}  disk throughout. This arises as a result of the gas velocity dispersion remaining constant or decreasing only slightly while the disk mass distribution declines exponentially as a function of radius.
\end{abstract}


\section{Introduction}
\label{intro}

In this paper we address the vertical distribution of neutral atomic hydrogen (\mbox{H\,{\sc i}}) in four dwarf irregular (dIrr) galaxies that are part of THINGS (The  \mbox{H\,{\sc i}} Nearby Galaxy Survey; \citealt{wal08}). These objects occupy the low mass, low luminosity tail of the galaxy luminosity function. They have accordingly low metallicities, in agreement with the finding that star formation in these systems occurs episodically, in bursts, with long periods during which apparently little star formation is taking place. Contrary to larger spiral galaxies, where stars dominate, in dwarf galaxies stars and \mbox{H\,{\sc i}} make up the baryonic mass content on an equal footing, with little or no H$_2$ being traced through CO observations \citep{tac87,tay98}. The jury is still out on the relation of CO to H$_2$ in low--metallicity, low--mass environments encountered in dwarf galaxies, although it seems clear that a low CO signal does not necessarily mean a correspondingly low H$_2$ mass \citep{wil95,bol08,ler09}. Their low metallicity and the fact that a large fraction of their gas has not yet been turned into stars is often used to argue that dwarf galaxies, within the paradigm of hierarchical galaxy formation, then must be broadly similar to the building blocks at high redshift that eventually formed today's larger galaxies. In accordance with models of hierarchical structure formation dwarf galaxies are dark matter dominated, something which is born out by observations \citep{car88,ash92,oh10}. They constitute the largest number of galaxies in the present day observable universe \citep{mat98}.

\citet{bri02} postulated that the gas scale height in dIrrs was larger, both in absolute as well as relative sense, than in larger, spiral galaxies. This has several implications, especially if indeed nearby dIrrs are broadly similar to those found at large look--back times. For example, their cross section for intercepting distant QSOs, giving rise to damped Lyman--$\alpha$ signatures, will be much larger. Also, a thicker gas layer will tend to allow \mbox{H\,{\sc i}} shells and bubbles to grow to larger diameters, preventing early break--out and thus reducing the likelihood that metals pollute the intergalactic medium.

The vertical stellar distribution has been extensively studied observationally \citep[see for a review of his early papers][]{vdk88}. The fact that the 
stellar distribution shows a moderate flaring at large radii was shown observationally by \citet{de97}, and predicted from a general self-consistent model by \citet{nar02}. In broad terms, the observed constant velocity dispersion of the gas combined with a decreasing mass surface density (as traced by the exponential decline of the stellar light distribution) predicts a flaring \mbox{H\,{\sc i}} layer. Early direct observations of this flaring were made in the Milky Way \citep{hen82} and M\,31 \citep{bri84}. 

Observing the \mbox{H\,{\sc i}} gas scale height directly is difficult, even for a perfectly edge--on galaxy, due to effects such as line--of--sight integration, beam smearing and optical depth effects as was already pointed out by \citet{san79}. More recent observations of the Milky Way are given by \citet{wou90} and \citet[][see also \citealt{kal09}]{lev06}, and for M\,31 by \citet{bra91}. Some other systems whose scale heights have been measured are NGC 3198 and NGC 2403 \citep{sic97}. No serious attempt has been made to date to investigate the vertical stellar and gas distributions in dIrrs. As discussed by \citet{bri02}, because their observed \mbox{H\,{\sc i}} velocity dispersion is not very different from that observed in spirals, but their mass surface density is considerably lower, dIrrs must have a larger scale height in \mbox{H\,{\sc i}} as compared to spiral galaxies in an absolute sense.

The origin of flaring of the gas layer in spiral galaxies has been treated recently in a self--consistently manner by \citet{nar02a}, who produced a numerical model of the vertical scale heights of the gas and stars in a gravitationally--coupled, three--component Galactic disk. In this paper, we extend this earlier work on the vertical stellar and gas distribution in spirals to dwarf galaxies, and determine the scale height of the gas as a function of radius analytically. We model the galaxy as a gravitationally coupled system of stars and gas subjected to the external force field of a dark matter halo, and use the model to obtain the \mbox{H\,{\sc i}} vertical scale height as a function of galacto--centric radius for four dwarf galaxies. In earlier papers we had assumed a flat rotation curve \citep{nar02,nar05}. This approximation is not warranted for dwarf galaxies which show a steadily rising rotation curve throughout most of their disk. Instead we explicitly take account here of a rising rotation curve, and show that the originally two--dimensional partial differential equations can still be written as one--dimensional or ordinary differential equations, as in the case for the giant spiral galaxies which have flat rotation curves. 

We will concentrate in this paper on the properties of the bulk of the gas and therefore ignore the contribution by the $\sim10$ percent of  ``extra--planar gas" found by Fraternali and collaborators  \citep{mar11,fra08}, most of which seems to be related to energy input from supernovae into the interstellar medium. We describe our methodology in Sects.~\ref{method} and \ref{num-calc}, justify the values for the input parameters to the model in Sect.~\ref{input}, and present our results and their discussion in Sects.~\ref{result}.

\section{Model}
\label{method}

\subsection{Gravitationally--coupled, two--component, galactic disk model}

We use a two--component galactic disk model of gravitationally--coupled stars and \mbox{H\,{\sc i}} gas subjected to the external force--field of a dark matter halo with a pseudo--isothermal density profile. In this model, the stars and gas are assumed to be present in the form of concentric, thin, axisymmetric disks embedded within each other \citep{nar02,ban07}.

The joint Poisson equation for an axisymmetric system of stars and gas is given by:

\begin{equation}
\frac{1}{R}\frac{\partial}{\partial R}(R\frac{\partial \phi_{total}}{\partial R}) +  \frac{\partial^2\phi_{total}}{\partial z^2} = 4\pi G(\sum_{i=1}^{2} \rho_i + \rho_{h})  
\end{equation}

where $\phi_{total}$ is the net potential of the disk due to the stars, the \mbox{H\,{\sc i}} gas and the dark matter halo, $\rho_i$ with $i = 1$ to 2 denotes the mass volume density for each of the disk components (stars, gas) while $\rho_h$ denotes the same for the halo, and is given by \citep{zee88}:

\begin{equation}
\rho_h(R,z) = \frac{\rho_0}{1+\frac{R^{2} + z^{2}}{R_c^{2}}} 
\end{equation}

where $\rho_0$ is the central core density of the halo, and $R_c$ is the core radius, as applicable for a pseudo--isothermal density distribution. One may note here that since dwarf galaxies have rotation curves that are generally rising out to the last measured point, the radial term in equation~(1) cannot be ignored as in the case for large spirals with almost flat rotation curves \citep{nar05,ban08}. 

The equation of hydrostatic equilibrium for the $i^{th}$ disk component in the $z$ direction is given by \citep{roh77}:

\begin{equation}
\frac{\partial }{\partial z}(\rho_{i}<(v_{z}^{2})_{i}>) + \rho_{i}\frac{\partial \phi_{total}}{\partial z} = 0
\end{equation}

where $<v_{z}^{2}>_{i}$ is the mean square velocity of the $i^{th}$ disk component. equation~(3) constitutes two equations, one each for the density of each of the two disk components. In order to obtain the vertical density distribution and hence the scale height of \mbox{H\,{\sc i}}
 at each $R$, one needs to solve equation~(1) and the set of two equations given by equation~{3} jointly, which together constitute a set of three coupled differential equations in the variables $\rho_{i}$ and $\phi_{total}$ in two dimensions $R,z$. Even if all the initial conditions are known, obtaining a rigorous solution of the above equations is going to be extremely complicated and computationally expensive. Thus we look for suitable approximations to simplify the problem. 

\subsection{Approximations}
\label{approx}
Combining equation~(1) and (3) and assuming isothermal condition for each of the two disk components we get:

\begin{equation}
<(v_{z}^{2})_{i}> \frac{\partial}{\partial z}[\frac{1}{\rho_{i}}\frac{\partial \rho_{i}}{\partial z}] = -4\pi G(\sum_{i=1}^{2} \rho_i + \rho_{h}) + 
\frac{1}{R}\frac{\partial }{\partial R}(R\frac{\partial \phi_{total}}{\partial R})
\end{equation}

Now, we replace the radial derivative term by its approximate value by utilising our knowledge of the shape of the observed rotation curve, which will effectively reduce the above equation to a one dimensional problem along the $z$ direction, as follows:

\begin{equation}
 { {(R\frac{\partial \phi_{total}}{\partial R})}_{R,z} } =  {{({v_{rot}}^{2})}_{R,z}}
\end{equation}

where $ { {(v_{rot})}_{R,z} } $ is the intrinsic rotational velocity at any $R,z$. However, we do not {\em a priori} know  $ {({v_{rot}})_{R,z}} $ as a function of $z$ at a given $R$. However, at each $R$, the observed rotation curve gives the intensity--weighted average of $ {({v_{rot}})_{R,z}} $  along the vertical direction. We approximate the value of $ {({v_{rot}})_{R,z}} $ at each $z$ by the commonly used intensity--weighted average velocity, i.e., the rotational velocity given by the observed rotation curve. As a result of this approximation, equation~(4) reduces to

\begin{equation}
<(v_{z}^{2})_{i}> \frac{\partial}{\partial z}[\frac{1}{\rho_{i}}\frac{\partial \rho_{i}}{\partial z}] = -4\pi G(\sum_{i=1}^{2} \rho_i + \rho_{h}) + 
\frac{1}{R}\frac{\partial }{\partial R}{({{v_{rot}}^{2}(R))}_{obs}}
\end{equation}

The last term on the r.h.s now becomes a function of $R$ only, and can be easily determined from the gradient of the observed rotation curve. We note that the presence of the last term on the r.h.s dilutes the effect of the net local mass density as far as the downward gravitational pull is concerned. Simple calculations show that in the rising part of the rotation curve the value of the term is of the same order of magnitude, but entering with opposite sign, as that of the net local gravity which results in a more puffed--up gas distribution and hence a larger value for the gas scale height. 

We further stress that in solving equation (6), we have used the dark matter core density ($\rho_{h}$) and the last term involving the gradient of the rotational velocity, both of which have been obtained from the observed rotation curve. While the latter has been determined directly from the
observed curve, $\rho_{h}$ was estimated after subtracting the disk contribution to the rotational velocity based on the calculated mass model of the
disk.

\section {Numerical calculations}
\label{num-calc}
The two coupled, second--order, ordinary differential equations given by equation~(6) are solved numerically using the Fourth order Runge--Kutta Method of integration in an iterative fashion \citep[see][]{nar02,ban07}. Since each of the two equations is a second--order differential equation, we need to state the following two initial conditions at the mid--plane (i.e., $z = 0$) for each component for solving the equations:

\begin{equation}
\rho_i = (\rho_0)_i  \qquad \frac{d\rho_i}{dz} = 0
\end{equation}

However, the modified mid--plane density $(\rho_0)_i $ for each component is not known {\em a priori}. Instead the net surface density $\Sigma_i(R)$, given by twice the area under the curve of $\rho_i(z)$ versus $z$, known observationally, is used as a secondary boundary condition. The required value of $(\rho_i)_0$ is then determined by trial and error, and eventually the $\rho_i(z)$ distribution is fixed. 

\section {Input Parameters}
\label{input}

As mentioned above, we have to provide as inputs the surface density $\Sigma(R)$ and the vertical velocity dispersion ${(\sigma_{z})}_{i} = <(v_{z}^{2})_{i}> ^{1/2}$ of each of the two components (stars and \mbox{H\,{\sc i}}) in order to solve equation~(6) at a given radius $R$, in addition to the observed rotation curve of the galaxy. Most of the input was taken from the compilation of THINGS \citep{wal08} papers. The stellar surface densities $\Sigma_*(R)$ are taken from \citet{ler08} for all dwarfs except NGC\,2366 which is taken from \citet{oh08}. Since a direct measurement of the velocity dispersion of stars ($\sigma_z^*$) is not available, it is calculated \citep[as described in][Appendix B]{ler08} based on the scale length of the disk $R_D$ and the surface density of the stars $\Sigma_*(R)$. This assumes a stellar disk in vertical hydrostatic equilibrium, neglecting gas gravity and thus slightly underestimates the stellar velocity dispersion. However, the response of the gas to the net galactic potential is governed by the value of its own velocity dispersion (see equation (6)), and so a small change in the vertical density distribution of the stellar component can change the calculated gas scale heights but marginally. The stellar disk scale lengths $R_D$ are again from \citet{ler08}. We furthermore need the scale height of the stars ($h_*$) in order to calculate the volume density,  $(\rho_0)_* $. We used $h_*$ = $R_D$/7.3 and an average flattening ratio \citep{kre02}. 

 We derived the \mbox{H\,{\sc i}} surface density $\Sigma_{HI}(R)$ and adopted a constant velocity dispersion ($\sigma_z^{HI}$) based on maps from \citet{wal08}. The resulting \mbox{H\,{\sc i}} scale height depends strongly on the assumed value of the vertical velocity dispersion $\sigma_z^{HI}$, which lies between 7 -- 9 km\,s$^{-1}$. Our calculations show that a change in the value of the assumed \mbox{H\,{\sc i}} velocity dispersion by 1 km\,s$^{-1}$ (or approximately 10 percent) changes the \mbox{H\,{\sc i}} scale height by 14--15 percent. Therefore, the accuracy of the determined \mbox{H\,{\sc i}} scale height depends on the accuracy with which $\sigma_z^{HI}$ can be determined. \citet{tam09}, using the same THINGS data, published a more complete analysis of the velocity dispersions. Most galaxies show a gradual fall in \mbox{H\,{\sc i}} velocity dispersion with radius, $\sim $ a few km s$^{-1}$ per unit of R$_{25}$. In addition to assuming a constant value, we use a smoothly falling velocity dispersion to model those two galaxies of our sample, Ho\,II and IC\,2574, for which this more complete velocity dispersion information is available \citep{tam09}.

The dark matter central core densities $\rho_{0}$ and core radii $R_c$ are taken from \citet{blo08} except for Ho\,II where we used \citet{bur02}. These are determined from the observed rotation curves assuming rotational equilibrium, and a dark matter distribution modelled as a pseudo--isothermal halo.  The observed rotation curves are taken from \citet{blo08} for DDO\,154 and NGC\,2366, from \citet{oh08} for IC\,2574 and from \citet{bur02} for Ho\,II. Key parameters used as input are listed in Table~\ref{table1}. We note here that de Blok et al. (2008) estimated the disk contribution to the rotation curve by employing mass models for the stellar and the gaseous disks. This was subtracted from the observed rotation curve to obtain the contribution of the dark matter halo from which its core density and core radius are derived. The mass model of the disk is well constrained and allows for small variations in its parameters only that in turn may result in small changes in the deduced halo parameters; the effect on the gas scale height will be minimal.

The essence of the radial mass distribution of the galaxy is effectively captured in the rotation curves. Dwarf galaxies in general have a rising rotation curve, i.e., the rotation velocity of these galaxies increases steadily with radius as far as 3--4 $R_D$, unlike spiral galaxies which sport flat rotation curves. 
We further note that in our previous work, we had modelled the observed gas scale height data to obtain the best--fit halo parameters, as was done for the Galaxy \citep{nar05}, M31 \citep{ban08} and UGC 7321 \citep{ban10}. Here the inverse approach is used, namely the observed rotation curve data is modelled to obtain the dark matter halo parameters and then the resulting \mbox{H\,{\sc i}} vertical scale height distribution is derived. This approach is valid as long as the halo profile can be approximated by a pseudo--isothermal sphere.

\section{Results and Discussion}
\label{result}

\begin{figure}
\begin{center}
\includegraphics[width=80mm,trim=0mm 0mm 5mm 0mm,clip]{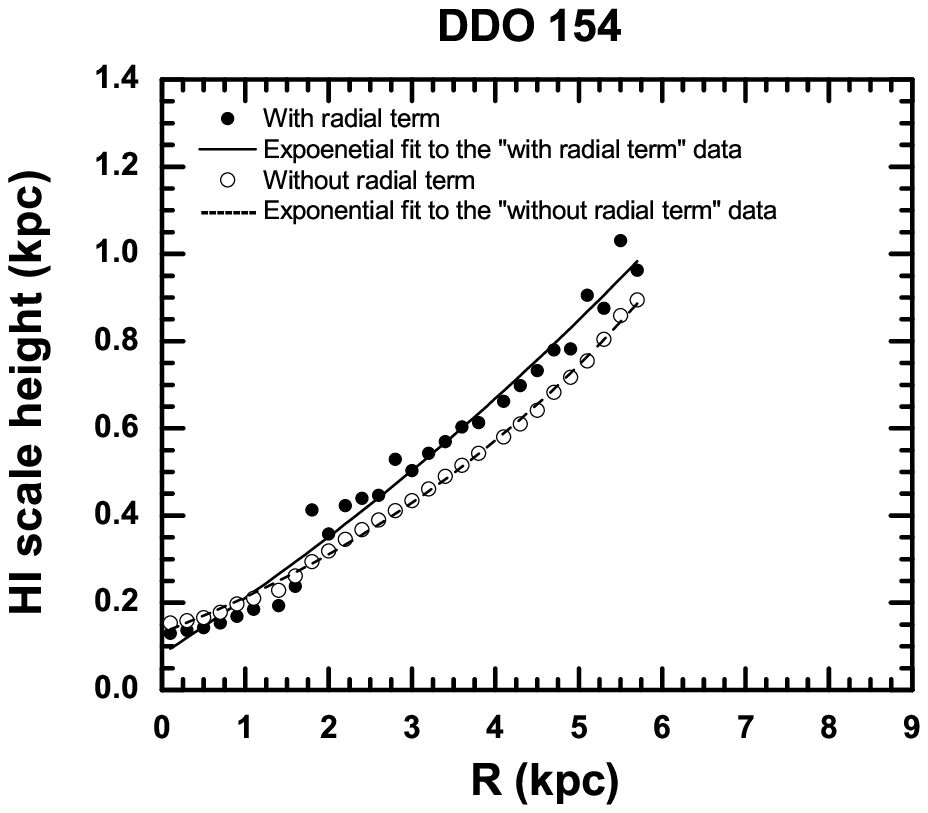}
\includegraphics[width=80mm]{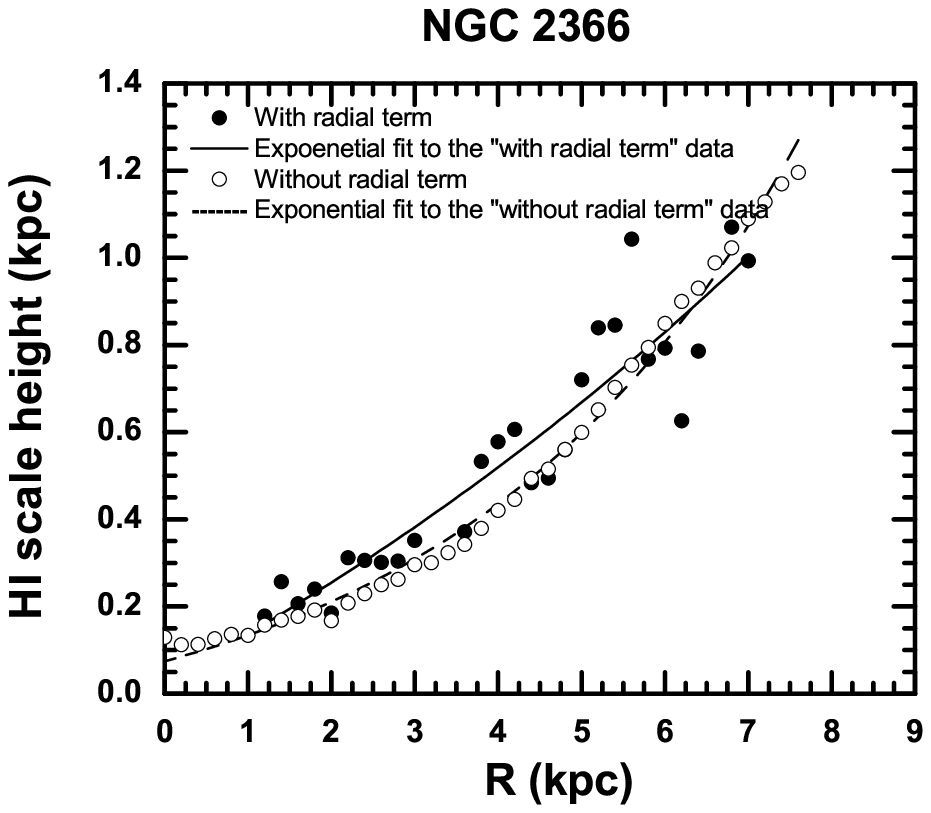}
\includegraphics[width=80mm]{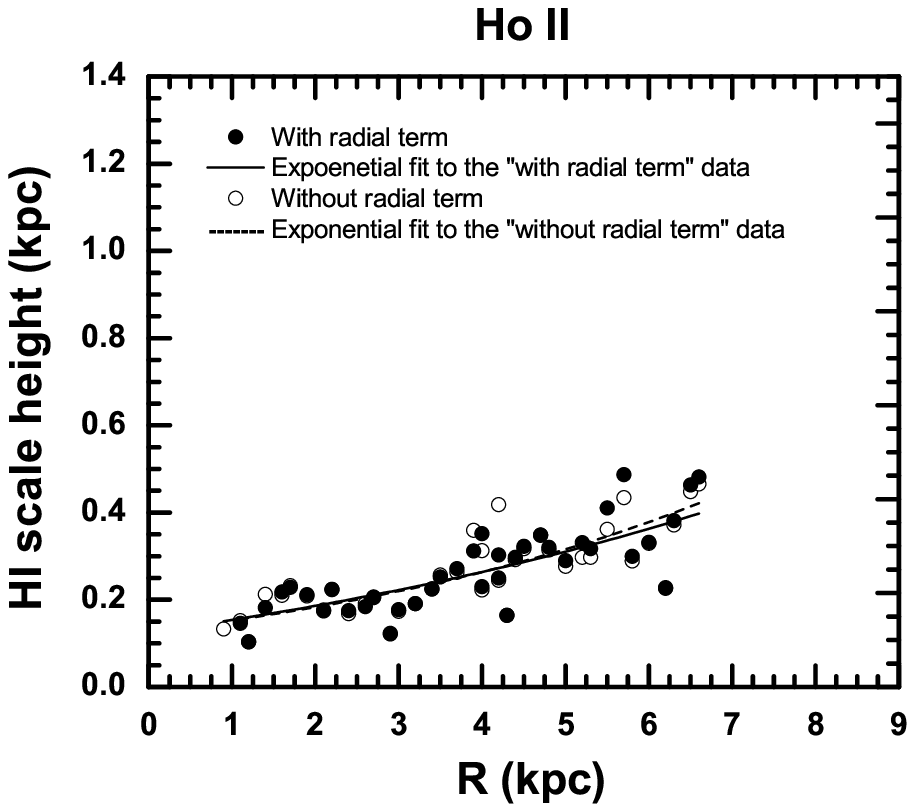}
\includegraphics[width=80mm]{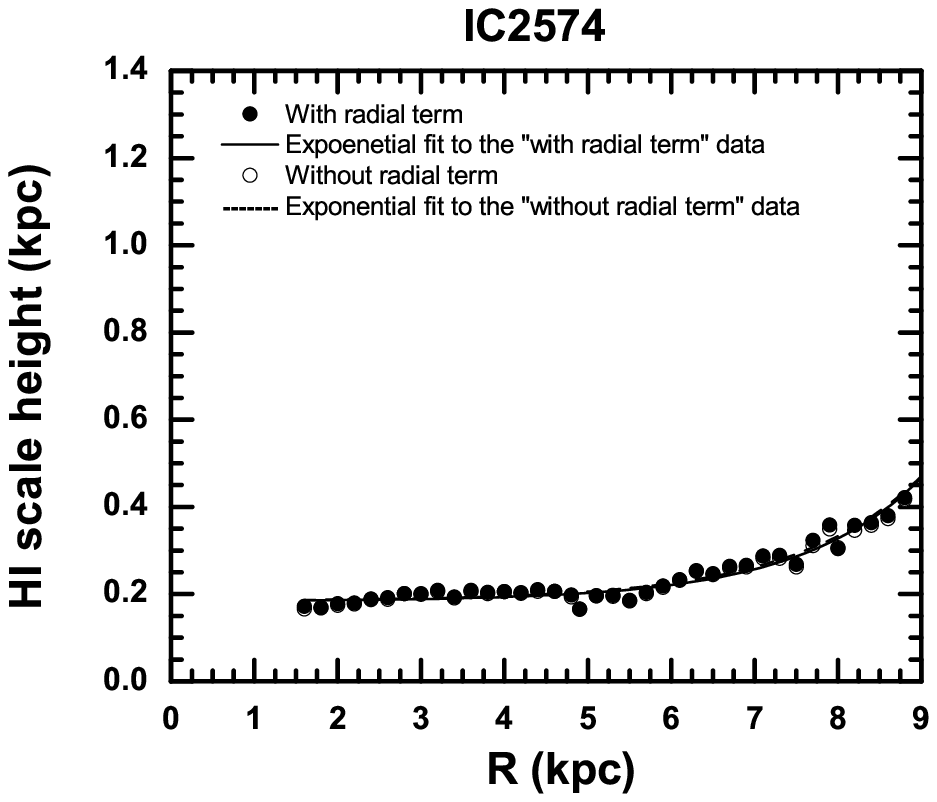}
\hspace{3mm}
\includegraphics[width=80mm]{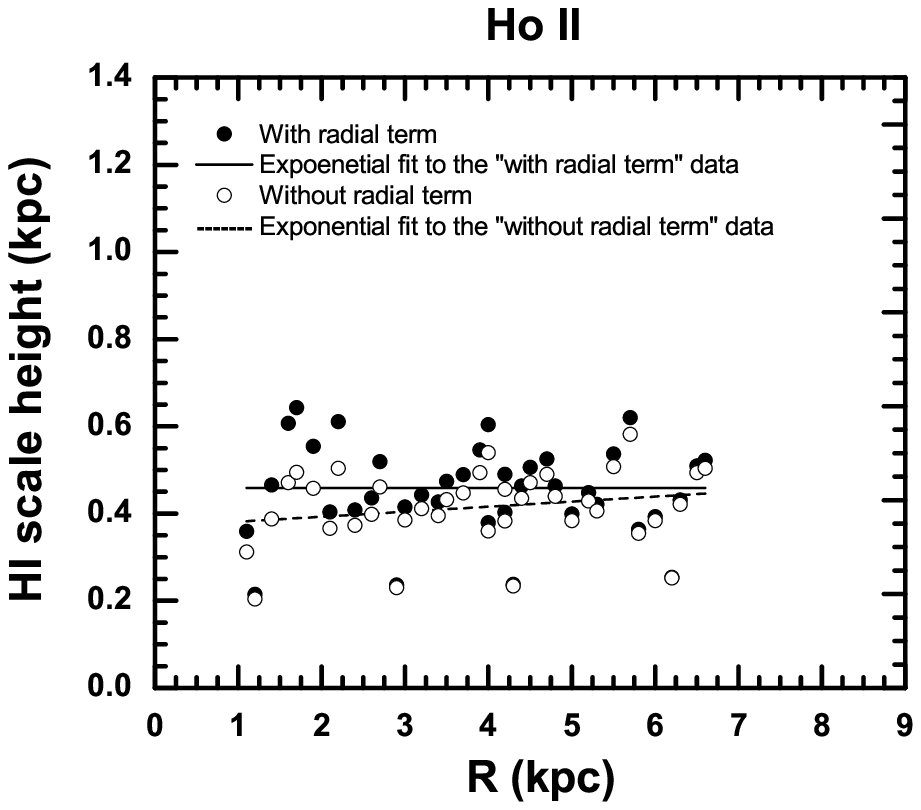}
\includegraphics[width=80mm]{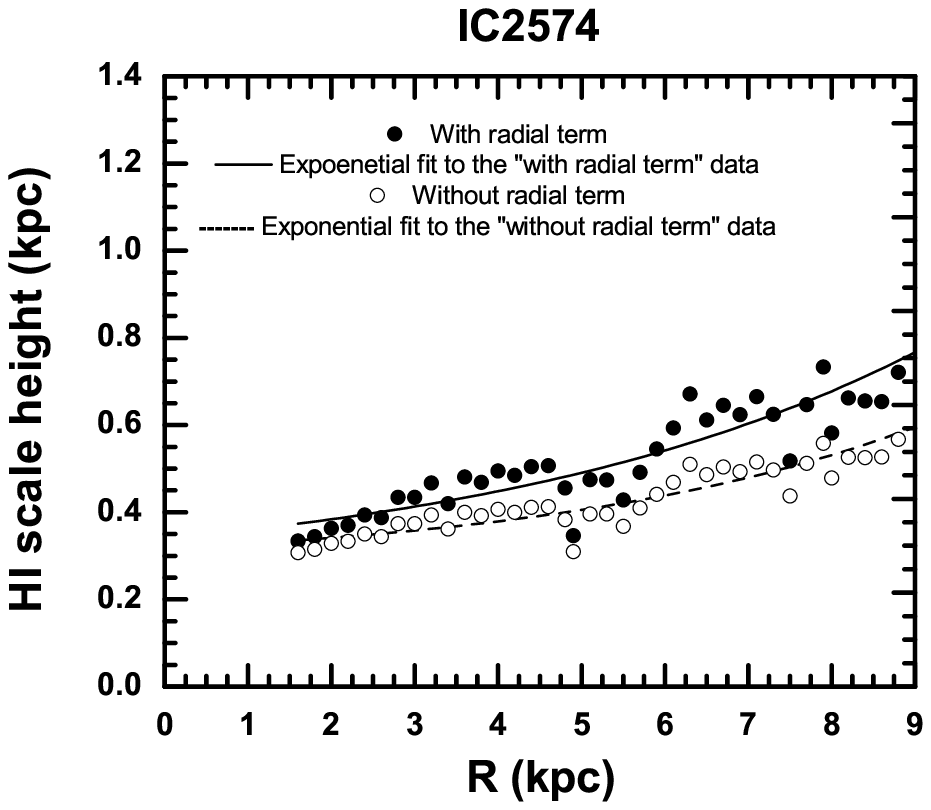}
\hspace{3mm}
\caption{\mbox{H\,{\sc i}} scale height (HWHM) versus $R$ for the four galaxies analysed here, with and without taking into account the effect of the radial term in the Poisson equation (see legend in each of the panels). The drawn lines are exponential fits to the calculated scale heights. The top four panels show our calculations, assuming for each galaxy a constant velocity dispersion as a function of radius. As explained in the text, the inclusion of the radial term in the Poisson equation which takes care of the rising rotation curve, results in a slghtly larger scale height in general. The two bottom panels show the inferred \mbox{H\,{\sc i}} scale height for Ho\,II and IC\,2574 when using a gradually decreasing velocity dispersion instead (see text for details).}
{\label{fig1}}
\end{center}
\end{figure}

We present in the top four panels of Figure~\ref{fig1} the resulting \mbox{H\,{\sc i}} scale height versus radius for DDO\,154, Ho\,II, IC\,2574 and NGC\,2366, assuming for each galaxy the constant \mbox{H\,{\sc i}} velocity dispersion as given in Table~\ref{table1}. For each galaxy, the scale height calculations were done twice: first by neglecting the radial term in the Poisson equation, mimicking the approach used for spiral galaxies \citep{nar02,ban07}, and a second time including the radial term. The two results differ from each other by about 10--20 percent, indicating that the exact shape of the rotation curve only enters as small, systematic correction. As mentioned above, the effect of the radial term is to somewhat counteract net local gravity, resulting in a larger value for the gas scale height. 

The individual data points follow, with some dispersion, a smooth increase with growing radius. Most of this dispersion is due to radial variations in the input parameters and associated uncertainties, such as when calculating the derivative of the rotation curve. The combined effect of these uncertainties is of the order of 100\,pc.

A parameter which has a more substantial influence  on the result is the velocity dispersion, and its dependence on radius. The two bottom panels of Figure~\ref{fig1} show for two of the targets, Ho\,II and IC\,2574, the effect of using a gradually declining velocity dispersion rather than a constant value. The rapid increase in \mbox{H\,{\sc i}} scale height for IC\,2574 becomes less pronounced as a result of the higher velocity dispersion near the centre, increasing the gas scale height there by about a factor of two, and a lower velocity dispersion near the last measured point, resulting in a relative increase across the gas disk of about a factor of two, rather than a factor of three earlier. The effect is even more pronounced for Ho\,II where we find an almost constant, but thick \mbox{H\,{\sc i}} disk throughout. This exercise illustrates the extent to which the resulting scale heights depend on accurate determinations of the velocity dispersion.

\begin{figure}
\includegraphics[width=80mm]{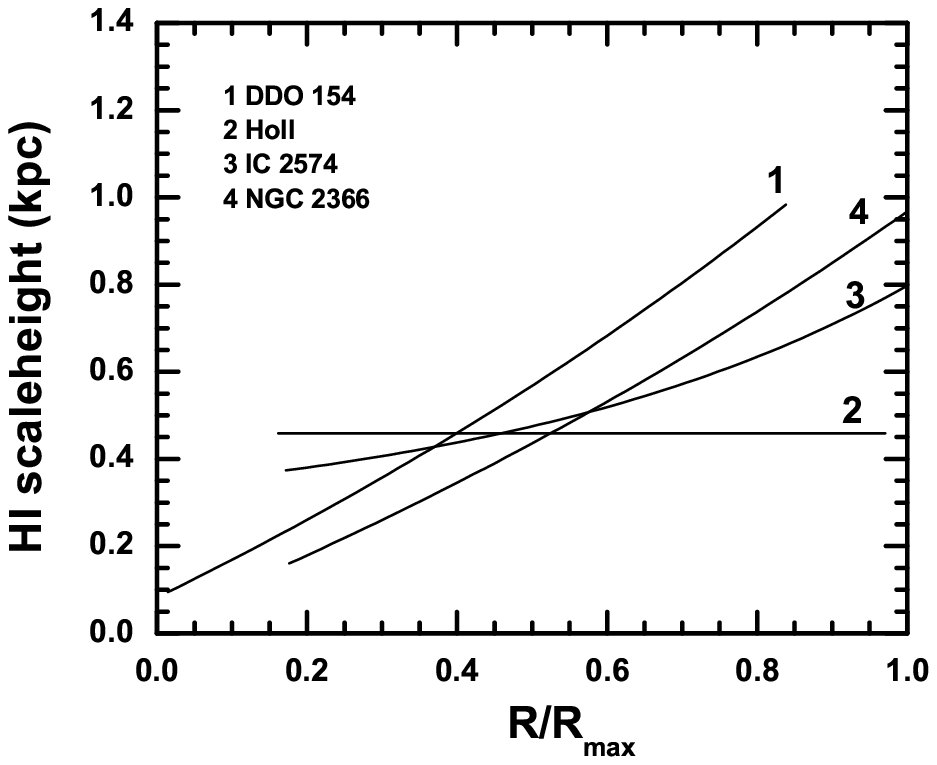}
\includegraphics[width=80mm]{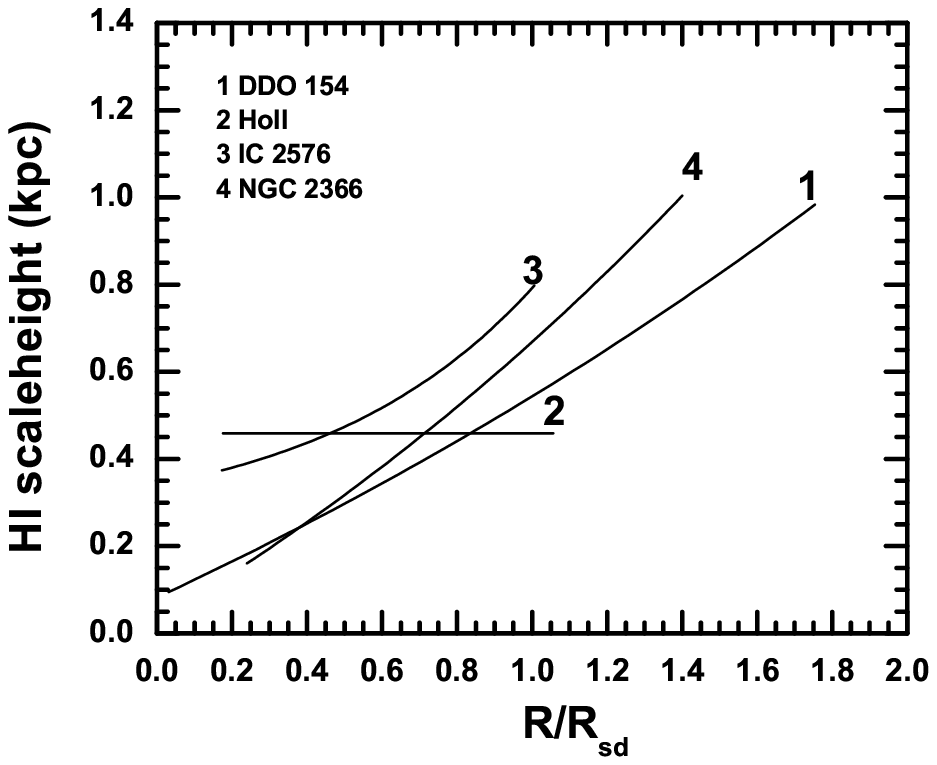}
\caption{Plots showing the \mbox{H\,{\sc i}} scale height (HWHM) versus normalised radius ($R/R_\mathrm{max}$ left--hand panel; $R/R_\mathrm{sd}$  right--hand panel) for the galaxies, taking into account the effect of the radial term in the Poisson equation and for Ho\,II and IC\,2574 using the radially decreasing velocity dispersion for the gas. Normalisation by $R_\mathrm{max}$ is with respect to the outermost extent of the \mbox{H\,{\sc i}} disk (from \citealt{bag10}), whereas normalisation by  $R_\mathrm{sd}$ is the radius where the mass surface density of gas plus stars reaches 4 M$_\odot$\,pc$^{-2}$. For clarity, rather than the individual radial values, the exponential fits to the analytical solution are shown.}
 {\label{fig2}}
\end{figure}

In the remainder of this paper, we will use for all galaxies the analytical solution to the gas scale height while taking the radial term in the Poisson equation into account. In the case of Ho\,II and IC\,2574 we take the the solution for the gradually decreasing velocity dispersion of the gas. Figure~\ref{fig2} then shows a pair of plots comparing the \mbox{H\,{\sc i}} scale height distribution as a function of normalised radius. We use two normalisations, the first one by maximum radial extent of the \mbox{H\,{\sc i}} disc, $R_\mathrm{max}$ \citep[taken from][]{bag10}. We also normalise by $R_\mathrm{sd}$, which corresponds to the radius where the mass surface density drops below 4\,M$_\odot$\,pc$^{-2}$. The $R_\mathrm{max}$ and the $R_\mathrm{sd}$ values for the different galaxies have been listed in Table~\ref{table1}.

We note that, with the exception of Ho\,II which has a thick gas disk throughout, the dwarfs all show \mbox{H\,{\sc i}} flaring. The rate of flaring is highest in NGC\,2366 and DDO\,154 which can be ascribed to these two galaxies having a more centrally concentrated dark matter halo than the other two systems.

\begin{figure}
\includegraphics[width=80mm,trim=0mm 0mm 5mm 0mm,clip]{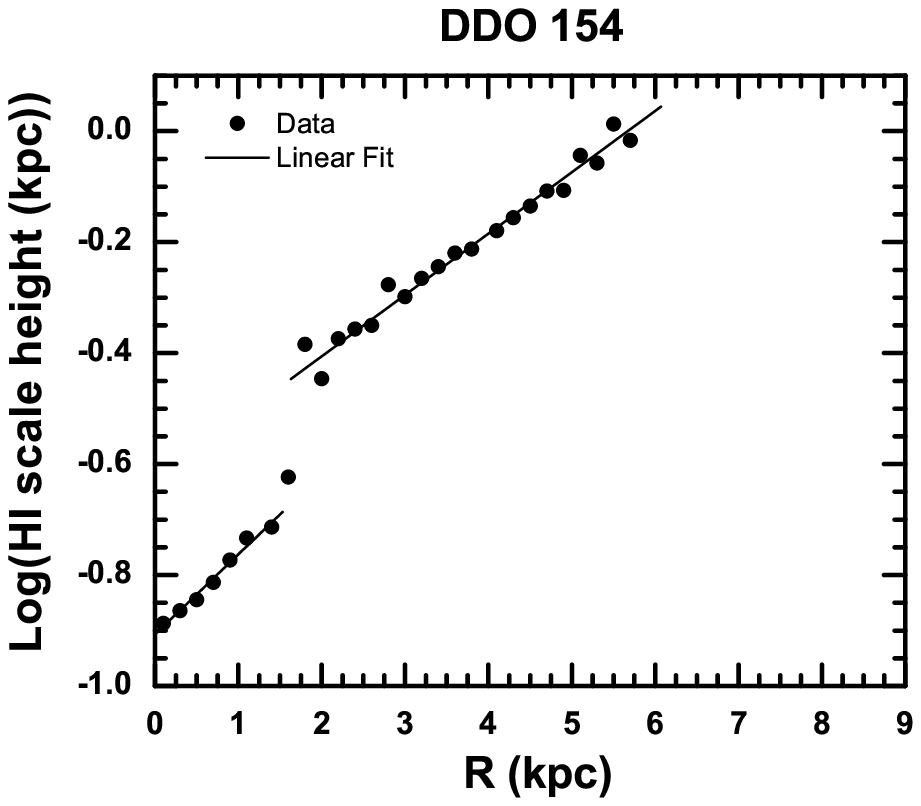}
\includegraphics[width=74mm,trim=5mm 0mm 0mm 0mm,clip]{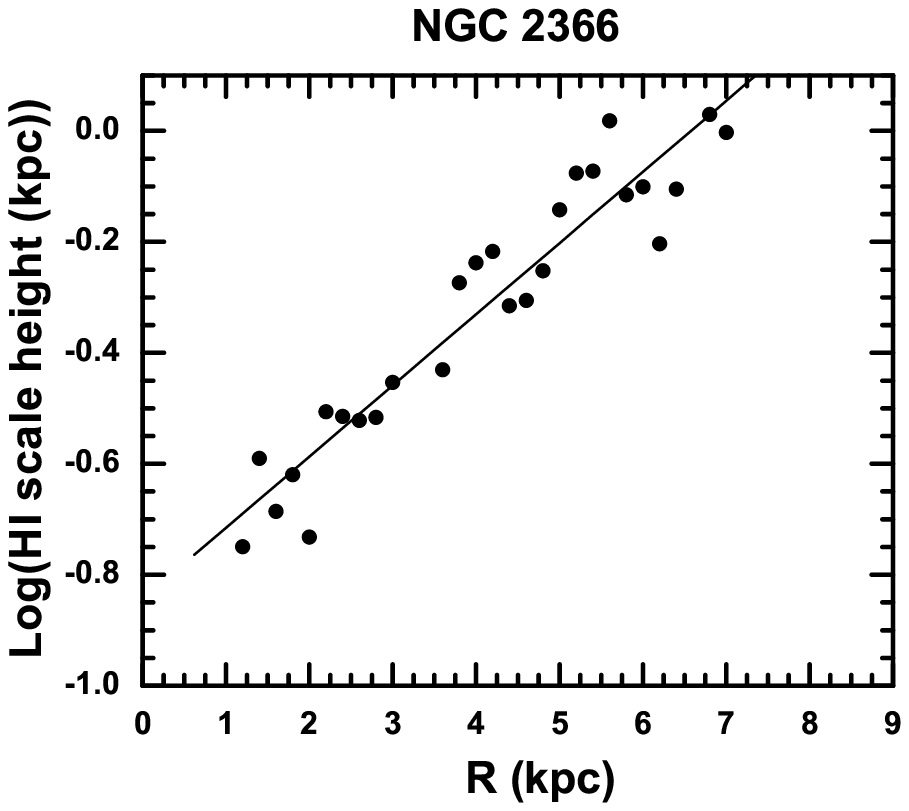}
\includegraphics[width=80mm]{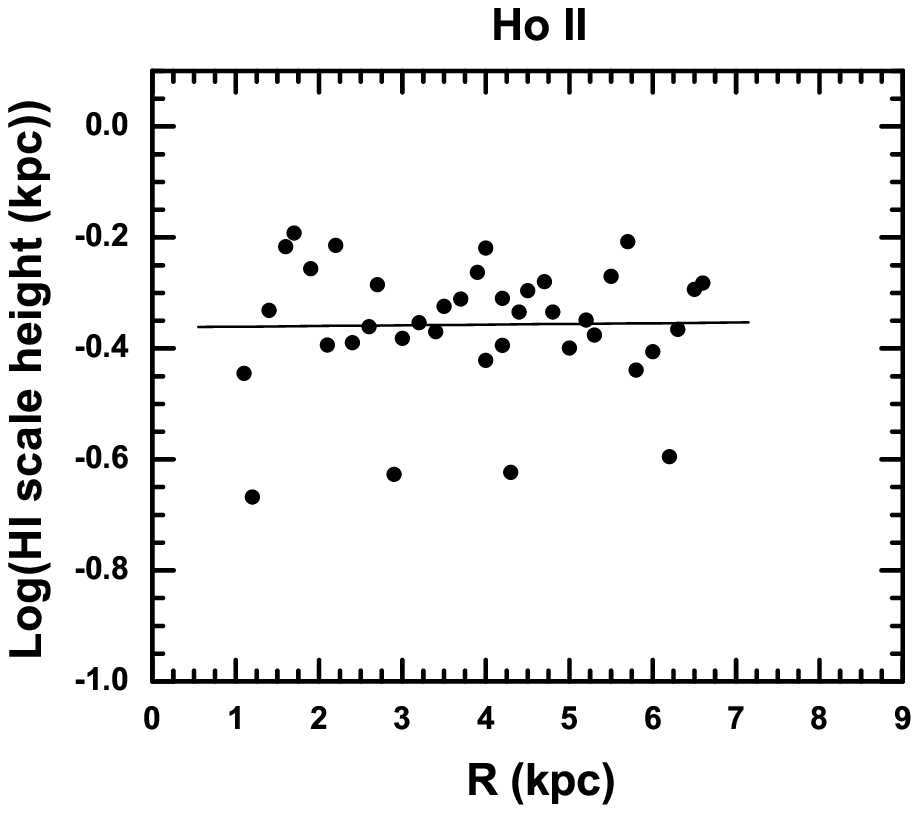}
\hspace{5mm}
\includegraphics[width=80mm]{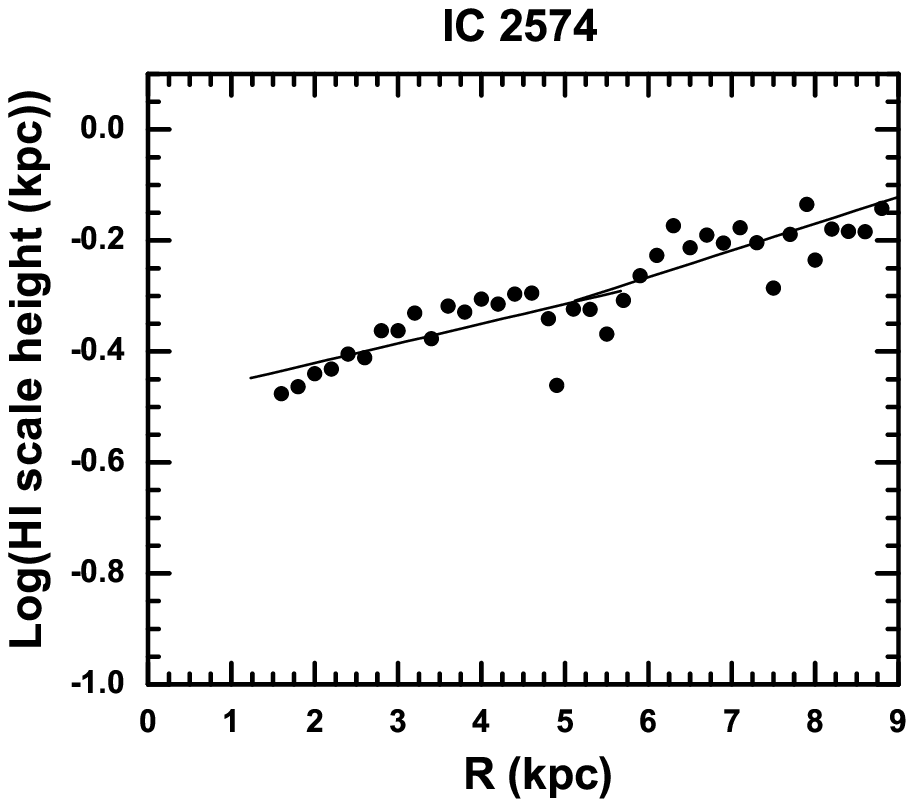}
\caption{Plots of logarithm of the \mbox{H\,{\sc i}} scale height versus radius, showing the scale height increases exponentially with radius for all the galaxies, but in some objects with a different scale length over different radial ranges (DDO\,154, and to a lesser extent IC\,2574).} 
 {\label{fig3}}
\end{figure}

\begin{figure}
\includegraphics[width=80mm]{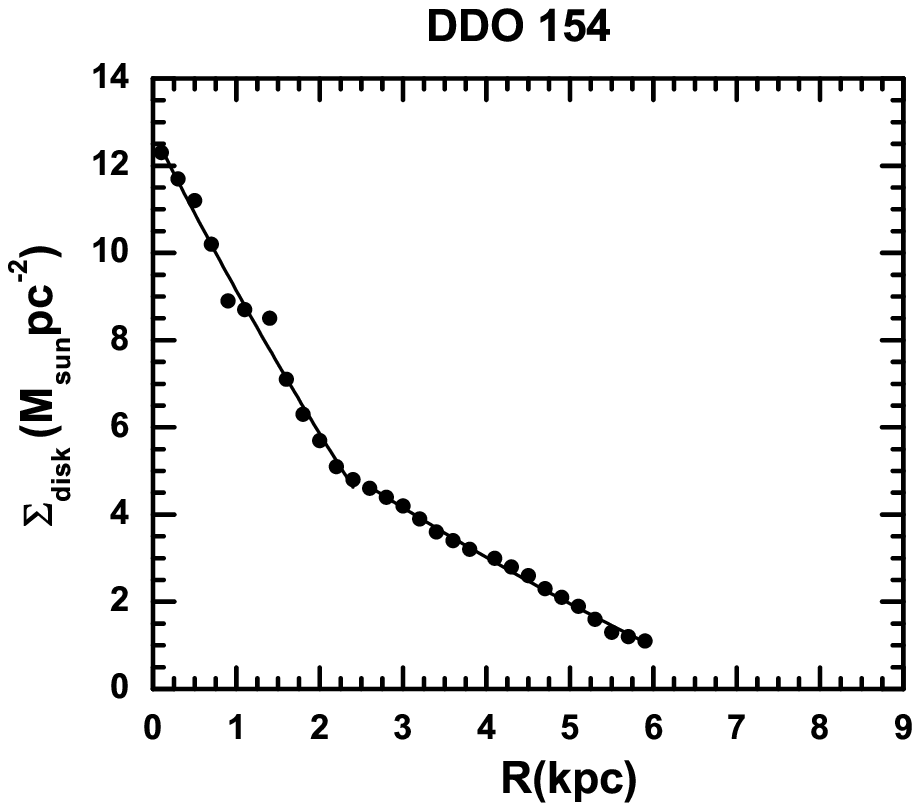}
\includegraphics[width=80mm]{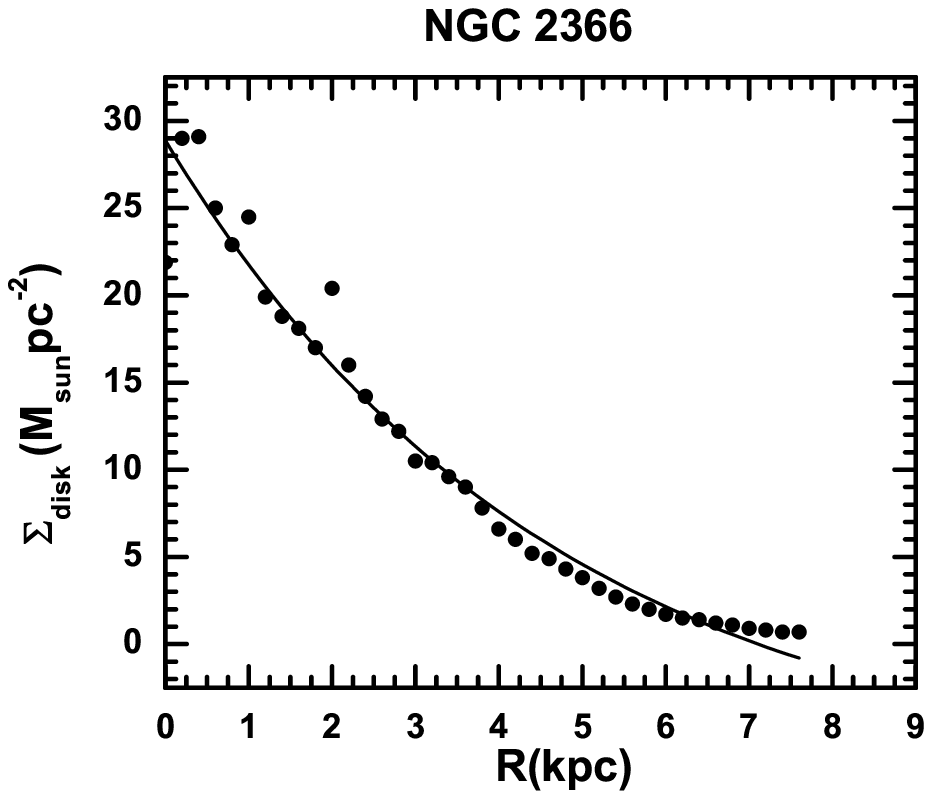}
\includegraphics[width=80mm]{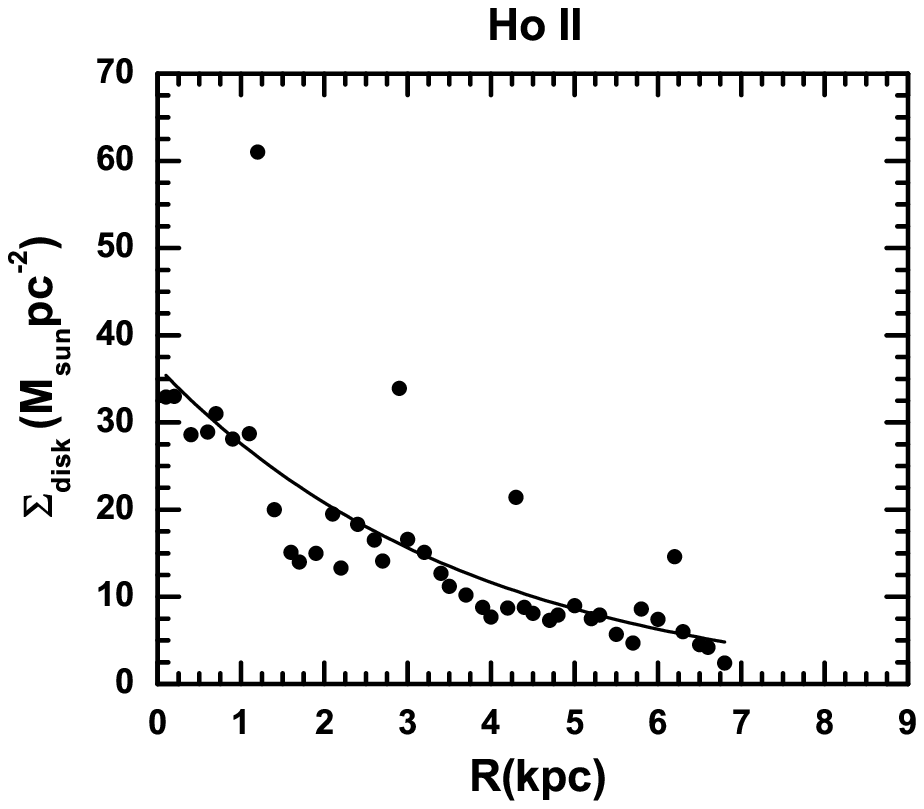}
\includegraphics[width=80mm]{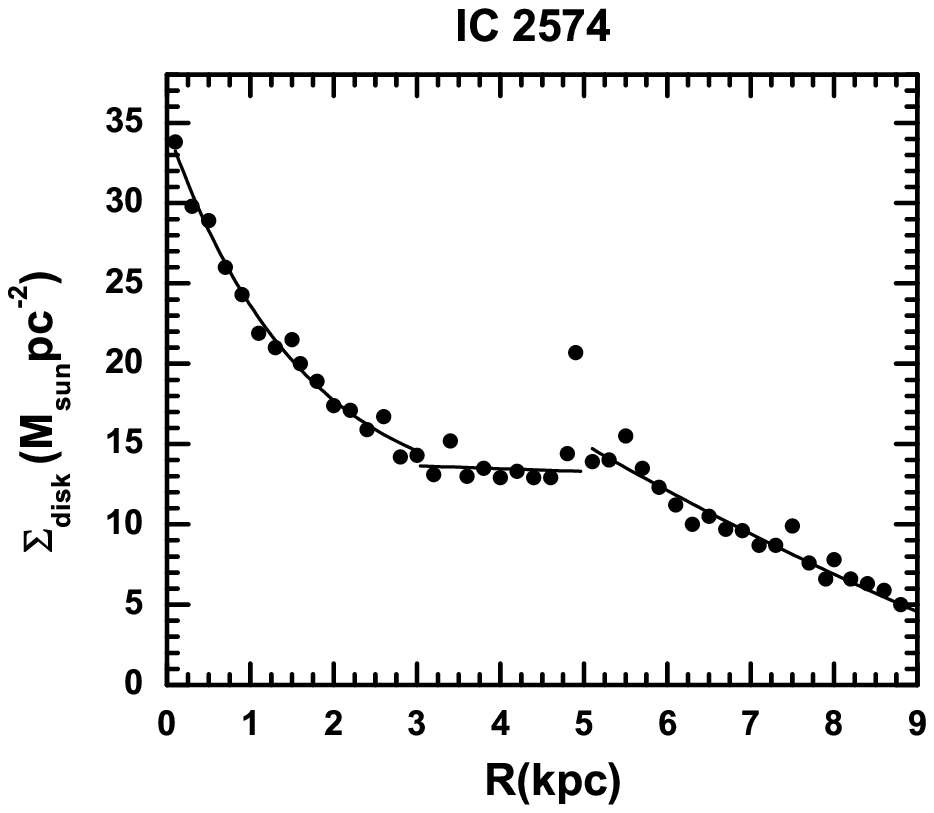}
\caption{Plots of the disk surface density (stars plus gas) versus radius for all galaxies. Comparison with Figure~\ref{fig3} confirms that for each galaxy, disk surface density plays a major role in regulating the gas scale height distribution.} 
{\label{fig4}}
\end{figure}

Figure~\ref{fig3} presents a collection of plots of the logarithm of the scale height versus radius, showing a near exponential behaviour, i.e. a linear slope, in each case. The curves for DDO\,154 and IC\,2574 display a break, necessitating two different slopes across two radial regimes, i.e., the gas scale height increases exponentially with radius with different scale lengths across the two regimes. It is interesting to compare this figure with Figure~\ref{fig4} which shows the disk surface density of gas and stars combined versus radius. Knees in the graphs of DDO\,154 and IC\,2574 occur at the same radius as the breaks in the exponential curves in Fig.~\ref{fig3}, confirming that the gas scale height depends strongly on the radial disk surface density distribution. 

This should make us pause for a moment as in the  adopted radial surface density distribution distribution we have ignored any contribution by molecular gas. In the introduction we mentioned that dIrr galaxies are deficient in CO which is used as a proxy to trace molecular gas. In fact, in galaxies with metallicities below $[12+ \log\mathrm{O/H}] < 7.5$ no CO emission has been detected \citep{tay98}. We also mentioned in the introduction that a lack of CO does not necessarily imply a lack of molecular gas. If present, this molecular gas likely follows the stellar light distribution \citep{you95,big08,ler09}. Its effects on the flaring \mbox{H\,{\sc i}} beyond the stellar disks of the dwarfs will therefore be negligible. In the inner parts the molecular gas, if present, will contribute to the mass density, effectively reducing the scale height locally.

Despite some modest potential adjustment at small radii due to a likely contribution of molecular gas to the mass density, and the inherent uncertainty in the measured parameters used as input to our analytical model, the main conclusion from this work is our confirmation of the flaring of \mbox{H\,{\sc i}} in dwarf galaxies. We also confirm the prediction by \citet{bri02} that the scale height in dwarfs is larger in absolute measure than in spiral galaxies. In dIrrs, we find scale heights from 200--400\,pc in the inner regions to 600-1000\,pc out to the last measured point.

Measuring the \mbox{H\,{\sc i}} scale height in dwarf galaxies directly is fraught with difficulties, even in the relatively ideal case of an edge--on dIrr. The projected \mbox{H\,{\sc i}} distribution on the sky is a combination of a flaring gas layer seen edge--on and a possible warp. It becomes a non--trivial task to extract the intrinsic gas density and scale height distribution \citep{oll95,oll96,war88,roy10}. In the case of more face--on dwarfs,  a break in the spectral correlation function can be used to infer the scale height of the gas \citep{elm01,pad01} but although this has been shown to work in principle in the LMC, this method requires high spatial resolution, high sensitivity data which is beyond what can routinely be obtained with current instruments.

As an independent check on our results, we compared the derived radial distribution of the \mbox{H\,{\sc i}} scale height with the size distribution of type 3 \mbox{H\,{\sc i}} holes, i.e., roughly spherical cavities in the \mbox{H\,{\sc i}} layer created by the combined effects of stellar winds and supernovae in young stellar associations \citep[see][for details]{bag10}. These structures are expanding and eventually their radius can exceed the thickness of the layer they find themselves in, at which moment they would be classified as a type 2 or type 1 hole, a type 1 hole corresponding to a complete blow--out. The maximum size of the type 3 holes encountered is therefore a lower limit on the local thickness of the gas layer. A comparison of the size distribution of  \mbox{H\,{\sc i}} holes in the four sample galaxies reveals that of the 20 type 3 holes all of them have radii that are in agreement with them being still fully contained within the gas layer. We note that if we were to adopt for Ho\,II and IC\,2574 the models with constant velocity dispersion, 2 out of 8 holes would exceed the thickness of the gas layer in Ho\,II and 1 out of 5 in IC\,2574, lending confidence to our approach and choice of input parameters.

In addition to the uncertainties in the input parameters and how they propagate, we made a couple of assumptions that should be mentioned here. In our approach, the stellar scale height is assumed  to be constant in order to obtain the vertical dispersion (Sect.~\ref{input}); the latter is used as input parameter in our model. Conversely, if the observed values for the stellar dispersion were known, we could obtain the resulting stellar scale height distribution {\em ab initio}. For spiral galaxies, the stellar scale height is shown to flare by a factor of 2--3 within the optical radius \citep{nar02a} as was also shown observationally by \citet{de97}.

Also, the error incurred in replacing ${(v_{rot})}_{R,z}$ at each $z$ by $(v_{rot})_{obs}$, which is higher in general, as per the approximations made in Sect.~\ref{approx}, results in a slight overestimation of the scale height. However, the error can not be more than a few percent as the effect of including the radial term itself is found to be less than ten percent and observations have shown that $(v_{rot})_{obs}$ {R,z} varies by ten percent or so along $z$ \citep{kam93}. 

In our analysis we assumed a pseudo--isothermal halo. In principle, if the \mbox{H\,{\sc i}} scale height can be determined reliably in an independent manner, our method could be expanded to include as free parameter the flattening of the dark matter halo.

\begin{figure}
\includegraphics[width=80mm]{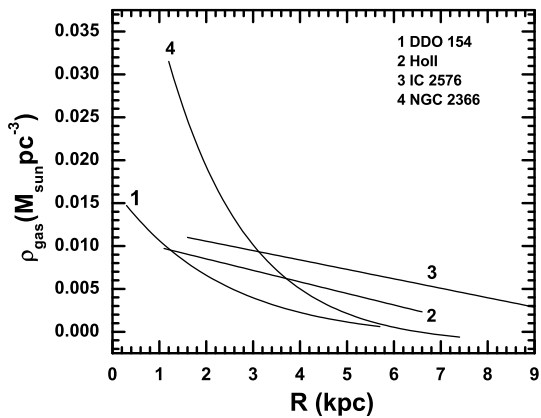}
\caption{A composite plot depicting the radial distribution of the mid--plane gas volume density for all galaxies.}
{\label{fig5}}
\end{figure}

Finally, in Figure~\ref{fig5} we show a composite plot for the radial distribution of the mid--plane gas volume density for the four galaxies based on the exponential fits to the relations in Fig.~\ref{fig1}. For reference, a volume density of 0.01\,M$_\odot$\,pc$^{-3}$ corresponds to 0.4\,H--atom\,cm$^{-3}$. The flaring of gas results in a lower mid--plane density. It would be interesting to investigate if this decrease in volume density explains the low star formation efficiency, a topic that will be dealt with as part of the LITTLE THINGS project \citep[see][for a description of LITTLE THINGS]{hun11}. 


\section{Conclusions}

In this paper we derive the \mbox{H\,{\sc i}} scale height as a function of radius for a group of four dwarf galaxies. We use a two--component model of gravitationally--coupled stars and gas in the force field of a dark matter halo to obtain the \mbox{H\,{\sc i}} scale height. We extended our earlier analysis by explicitly addressing the issue of the rising rotation curve of dwarf galaxies. Our approach gives more realistic results than earlier models in which the self--gravity of the gas was ignored. This is especially relevant for dwarf galaxies where stars and gas dominate the disk gravity, roughly at par with each other. We find that Ho\,II has a thick \mbox{H\,{\sc i}}  disk throughout; the other three dwarf galaxies studied in this paper show flaring \mbox{H\,{\sc i}} as a function of radius. Moreover, the overall \mbox{H\,{\sc i}} scale height of dwarfs is larger than in spiral galaxies. This is due to a combination of a lower mass density in dwarf discs as a function of radius for the same velocity dispersion as encountered in larger spiral galaxies. The flaring of \mbox{H\,{\sc i}} in the outer galaxy (beyond $R = 3-4 R_D$ ) is mainly controlled by the dark matter halo contribution of the net potential of the galaxy. The derived gas scale height is in agreement with the size distribution of type 3 \mbox{H\,{\sc i}} holes, i.e., roughly spherical, expanding cavities resulting from the energy deposited by rapidly evolving young stars. This type of holes is still contained within the gas layer and hence represent a lower limit to the gas scale height.\\ \\

\textbf{Acknowledgement:} We thank the anonymous referee for his valuable comments which has helped to improve the paper.

\newpage

\begin{table}
   \caption{Input parameters of the different galaxies.}
   \label{table1}
\begin{tabular}{lllllll}
\hline
Galaxy & ${\Sigma_*(0)}$ & ${R_{D}}$ & ${R_{25}}$ & ${\sigma_z^{HI}}$ & ${\rho_{0}}$ & ${R_{c}}$ \\
       & M$_{\odot}$\,pc$^{-2}$ & kpc & kpc & km\,s$^{-1}$ & M$_{\odot}$\,pc$^{-3}$ & kpc \\
&&&&&\\
\hline
DDO\,154  & 5.7  & 0.8 & 1.22 &8 & 0.028 & 1.34  \\
Ho\,II    & 27.8 & 1.2 & 3.26 &i) 7  & 0.009 & 10.0  \\
	  &      &     &      &ii)14. - 0.91R if R $<$ 3.3 & & \\
	  &      &     &      &   9.9 - 0.76(R - 3.3) otherwise & & \\
IC\,2574  & 24.6 & 2.1 & 7.53 &i) 7  & 0.004 & 7.23  \\
          &      &     &      &ii)10.8 if R $<$ 6.4 &  & \\
          &      &     &      &10.8 - 0.95(R - 6.44) otherwise &  &  \\
NGC\,2366 & 10.5 & 0.5 & 2.18 & 9 & 0.035 & 1.36 \\
\hline
\end{tabular}
\medskip 
The models for Ho\,II and IC\,2574 that were run for a constant gas velocity dispersion used ${\sigma_z^{HI}} = 7$\,km\,s$^{-1}$.
\end{table}

\clearpage
\newpage
\begin{table}
   \caption{$R_{max}$ (denoting the outermost extent of the  \mbox{H\,{\sc i}} disk) and $R_{sd}$ (the radius where the disk surface density falls to 4 M$_\odot$\,pc$^{-2}$) for the different galaxies}
   \label{table1}
\begin{tabular}{lll}
\hline
Galaxy & ${R_{max}}^{1}$ & ${R_{sd}}$ \\
       & $kpc$  & $kpc$ \\

\hline
DDO\,154  & 6.8 & 3.3 \\
Ho\,II    & 6.8 & 6.2 \\
IC\,2574  & 9.3 & 9.3 \\
NGC\,2366 & 6.8 & 5.0 \\
\hline
\footnote*{} The $R_{max}$ values have been taken from Bagetakos et al. (2010).
\end{tabular}

\end{table}


\clearpage
\newpage

\end{document}